\documentclass[fleqn,usenatbib]{mnras}

\usepackage{newtxtext,newtxmath}

\usepackage[T1]{fontenc}

\DeclareRobustCommand{\VAN}[3]{#2}
\let\VANthebibliography\thebibliography
\def\thebibliography{\DeclareRobustCommand{\VAN}[3]{##3}\VANthebibliography}

\usepackage{graphicx}	
\usepackage{amsmath}	
\DeclareMathOperator\erf{erf}

\newcommand{\pdf}{%
  f(x) = \frac{1}{x\sigma\sqrt{2\pi}} 
  \exp\left[ -\frac{1}{2}\left(\frac{\log{x}-\mu}{\sigma}\right)^{\!2}\,\right]
}

\title[Switchbacks and Turbulence]{The relation between magnetic switchbacks and turbulence in the inner heliosphere}

\author[A. Larosa et al.]{
A. Larosa,$^{1}$\thanks{E-mail:a.larosa@qmul.ac.uk}
C. H. K. Chen,$^{1}$
J. R. McIntyre$^{1}$
and V. K. Jagarlamudi$^{2}$
\\
$^{1}$Department of Physics and Astronomy, Queen Mary University of London, London E1 4NS, UK\\
$^{2}$Johns Hopkins University Applied Physics Laboratory, Laurel, MD, 20723, USA
}

\date{Accepted XXX. Received YYY; in original form ZZZ}

\pubyear{2023}

\begin{document}
\label{firstpage}
\pagerange{\pageref{firstpage}--\pageref{lastpage}}
\maketitle

\begin{abstract}
We investigate the relation between turbulence and magnetic field switchbacks in the inner heliosphere below 0.5 AU in a distance and scale dependent manner. 
The analysis is performed by studying the evolution of the magnetic field vector increments and the corresponding rotation distributions, which contain the switchbacks. We find that the rotation distributions evolve in a scale dependent fashion, having the same shape at small scales independent of the radial distance, contrary to at larger scales where the shape evolves with distance. 
The increments are shown to evolve towards a log-normal shape with increasing radial distance, even though the log-normal fit works quite well at all distances especially at small scales. The rotation distributions are shown to evolve towards the \cite{Zhdankin2012} rotation model moving away from the Sun. The magnetic switchbacks do not appear at any distance as a clear separate population. Our results suggest a scenario in which the evolution of the rotation distributions, including switchbacks, is primarily the result of the expansion driven growth of the fluctuations, which are reshaped into a log-normal distribution by the solar wind turbulence. 
\end{abstract}




\section{Introduction}\label{sec:intro}

\begin{figure*}
    \centering
	\includegraphics[width=\linewidth]{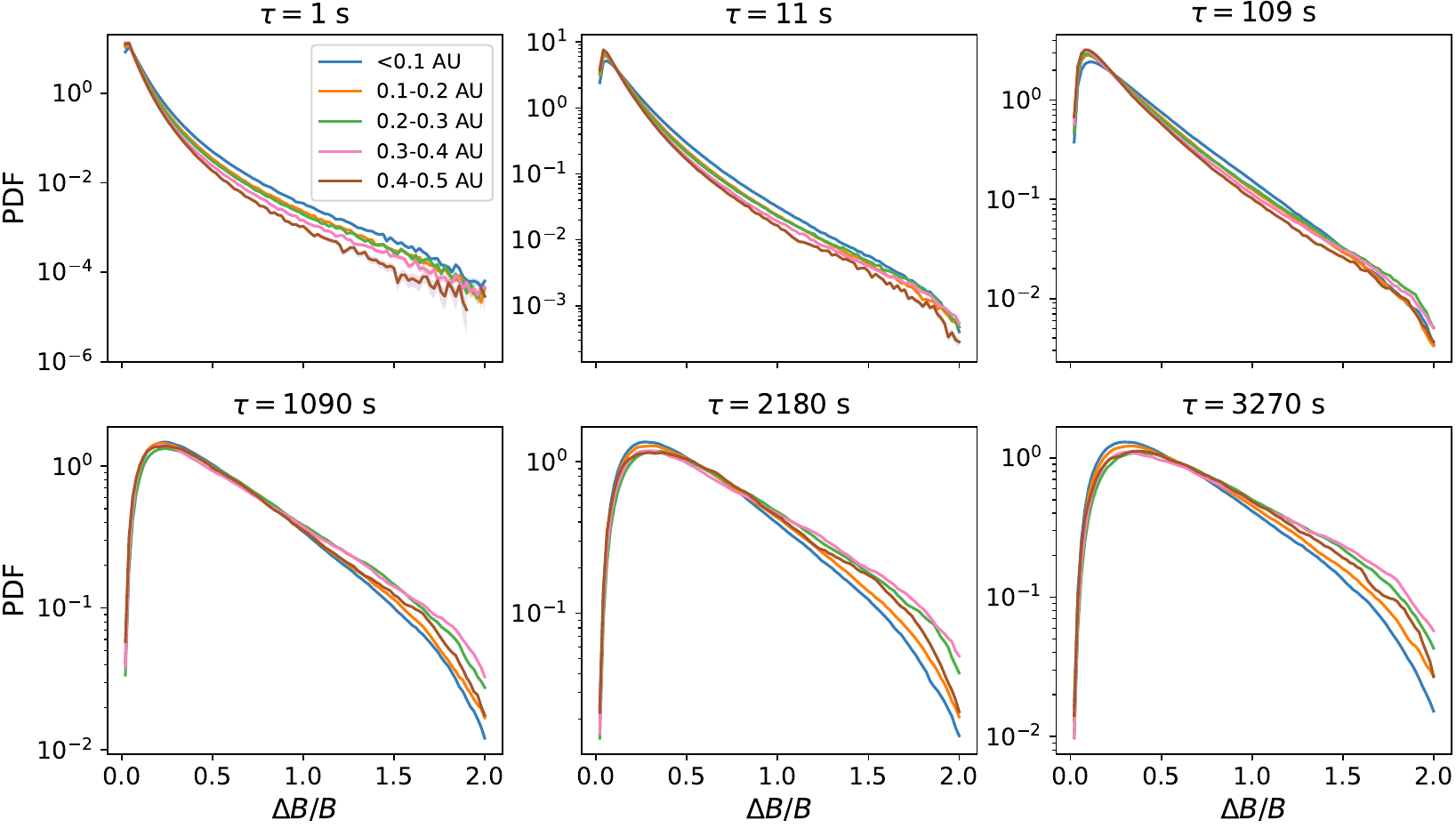} 
	\caption{Distributions of the magnetic field increments $\Delta B/ B$ : for each panel the increments are computed at different time scales, $\tau$, indicated. Different colors represent different heliocentric distances.}
	\label{fig:dBovBsameTau}
\end{figure*}

The solar wind is a turbulent medium whose properties evolve with radial distance from the Sun \citep{bruno_carbone2013}. Past works provided key findings such as the decrease of the power levels in the magnetic field magnitude and components \citep{Horbury_Balogh2001JGR_Helios_Ulysses}, the motion of the $1/f$ break \citep{bavassano_1_f_1982} and the ion spectral break \citep{bruno_trenchi2014_ionBreakRadDist} to lower frequencies with increasing radial distance, the transition from a more imbalanced spectrum of fluctuations to a more balanced one moving away from the Sun \citep{roberts1987Voyager, tu_marsch1995} (even at high latitudes \citep{breech_matt_2005}) and the steepening of the velocity spectrum from the Earth to 5 AU \citep{roberts2010HeliosVoyager}. 

In more recent years the Parker Solar Probe (PSP) \citep{FoxVelli2016} mission has improved our knowledge of the solar wind turbulence to distances below 0.3 AU. In PSP data the inertial range trace magnetic field spectrum was shown to evolve from a $-3/2$ slope to a $-5/3$ one with increasing radial distance \citep{chenTurb2020}. The controlling parameter of this evolution seems to be the cross-helicity \citep{JackCross2023}, therefore the $-3/2$ spectrum is associated with more imbalanced turbulence, consistent with previous observation at 1 AU \citep{podesta_borovsky_2010, chen2013ApJResEnergy, chen2016JPlPh..82f5302C}.
\cite{shi_velli2021} showed that the velocity field spectrum does not steepen with increasing radial distance up to 85 solar radii, also consistent with the 1 AU results, where $-3/2$ velocity spectra are seen at all levels of imbalance \citep{podesta_borovsky_2010, chen2013ApJResEnergy}.  
Regarding the $1/f$ range, \cite{ZesenHuang1ovf2023} and \cite{nooshin2023} have shown that below 0.3 AU the spectra can be shallower than $1/f$ and that they evolve towards $1/f$ with increasing advection time. The origin of this behaviour is still debated \citep{bill1986PhRvL..57..495M, Velli1989PhRvL..63.1807V, Verdini2012ApJ...750L..33V, Perez2013ApJ...776..124P, matteini2018ApJ, chandran2018JPl}.

In this evolving near-Sun turbulent medium PSP has revealed the presence of large amplitude highly Alfv\'enic magnetic deflections known as switchbacks (SBs) \citep{bale19, kasper19, DudokdeWitSBs2020, volodiaSBs2020ApJ...893...93K}. For most of these structures the magnitude of the field is constant \citep{larosa2021A&A}, therefore they can be thought as magnetic rotations to a good approximation. 
Several models have been proposed to explain their formation. As suggested in \cite{nourPSPreview2023} they can be grouped according to the invoked physical mechanism  as: reconnection, shear flows and Alfv\'en-wave/turbulence based models. 

In the reconnection driven models, a SB is formed either by a kink impressed on the newly open magnetic field line \citep{kasperSBs2020} or by the formation of a flux rope after an interchange reconnection process \citep{drakeSBs2021, agapitovSBs2022, Bale2023Natur.618..252B}. The propagation of the kink from the reconnection event location to PSP is shown to be possible when the non-linear and dissipative terms are discarded in the MHD equations by \cite{zankSBs2020ApJ}. Such a model is shown to be reliable when fitted to many SBs \citep{Liang2021SBsApJ}.
The reconnection driven models imply an ex-situ formation of the structures close to the Sun surface. Reconnection based models struggle to recover the Alfv\'enicity of the SBs, to explain the increasing occurrence rate with increasing radial distance \citep{teneraniSBs2021ApJ, pecoraSBs2022ApJ, Jagarlamudi_2023} and numerical results show that the kinks are unfolded before they can reach PSP \citep{wyper2022ApJ}. 

The shear driven models are based on the interaction of wind streams with different velocities. \cite{ruffoloSBs2020} propose that past the Alfv\'en surface the gradient in speed between adjacent flux tubes can exceed the Alfv\'en speed triggering the non-linear Kelvin-Helmholtz instability. The magnetic roll-up formed by this instability would appear as magnetic field reversals once crossed by PSP. In the model proposed by \cite{schwadronSBs2021} the shear that forms a switchback is due to the magnetic field footpoint motion from a region of slow solar wind to a region of fast wind. Due to this the fast wind overtakes the previously released slow wind creating a compression region and a folded magnetic field configuration. 
The shear driven models are consistent with, at a qualitative level, the increasing occurrence rate of SBs with radial distance but it is unclear whether they can reproduce the symmetric (leading vs trailing edge) shape of the velocity profile observed within SBs, if the compression they produce is as mild as observed for SBs \citep{larosa2021A&A} and if they can match the observed occurrence rate. 
\begin{figure*}
    \centering
	\includegraphics[width=\linewidth]{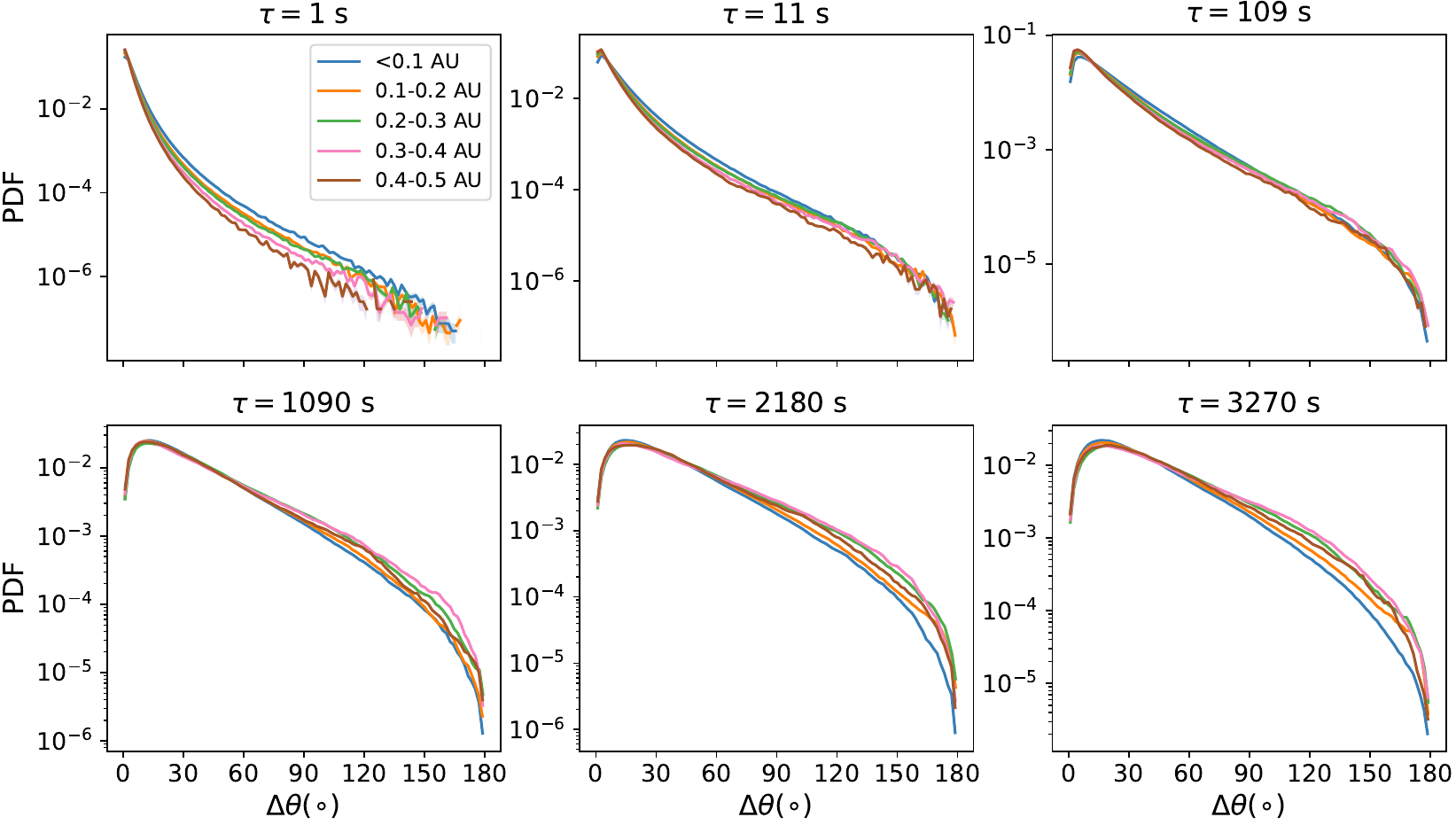} 
	\caption{Distributions of the magnetic deflection angles $\Delta \theta$. The order of the panels and the meaning of the colors is the same as in Figure~\ref{fig:dBovBsameTau}.}
	\label{fig:dthetasameTau}
\end{figure*}

The Alfv\'en wave/turbulence models are based on the fact that large amplitude Alfv\'en waves are an exact non-linear solution of the MHD equations if density, pressure and the magnitude of the magnetic field are constant \citep{barnesHollweg1974}. In these models SBs are then seen as spherically polarized Alfv\'en waves that have reached large amplitudes since $\delta B/B$ grows with the expansion.
This scenario has been explored in both numerical simulation \citep{squireSBs2020, shodaSBs2021, johnston2022SBs} and theoretical works \citep{MalletSBs2021, squire_mallet_spiral2022}. 
Expansion not only steepens the Alfv\'en waves, but it also acts, due to non-WKB effects, as a reflection term for large scales waves that enhance the turbulent development of the medium \citep[e.g.][]{grappin_velli1996JGR, velli1990CoPhC..59..153V, Cranmer2005ApJS..156..265C, Chandran2019JPlPh..85d9009C, squireSBs2020}, hence the name Alfv\'en wave/turbulence models.
These models, which are supported by many observations \citep{LiuSBs2023ApJ}, can reproduce most of the properties of SBs, but struggle in simulations to reproduce the filling factor observed in the data \citep{squireSBs2020, shodaSBs2021}. It is not clear, however, whether this is due to limited numerical resolution \citep{nourPSPreview2023}.

The different SB models are not mutually exclusive and a combination of them is possible. One could imagine, for example, that reconnection processes provide some of the seeds Alfv\'en waves that subsequently grow in amplitude in the expanding solar wind.

The interplay between SBs and turbulence is a matter of debate. From the comparison of the turbulence properties inside and outside the SB structures we know that SBs present: about one order of magntitude increase in power \citep{DudokdeWitSBs2020} that is more isotropically distributed between the parallel and perpendicular direction to the magnetic field \citep{saksheeSBs2022}, higher intermittency levels \citep{martinovicSBs2021}, higher residual energy \citep{bourouaineSBs2020}, a more developed inertial range \citep{DudokdeWitSBs2020}, a larger occurrence of small scales current sheet \citep{Jia2023arXiv230110374H} and enhanced kinetic Alfv\'en wave activity \citep{MalaspineKinALfSBs2022ApJ}. Despite these differences, both SBs and non SBs intervals present the same critical balance-like scaling in the inertial range \citep{saksheeSBs2022} and the ion spectral break at the same scale \citep{martinovicSBs2021}. 

An alternative way to investigate SBs, their link with turbulence and more in general the solar wind rotations is to study the full distribution of the magnetic field vector increments. This method has the advantage of being unbiased with respect to the choice of arbitrary deflection thresholds commonly used in the literature to define SBs \citep[e.g.][]{pecoraSBs2022ApJ} and to be more general since any different deflection threshold would correspond to looking at the right of different vertical cut in the rotation distributions. Furthermore, magnetic field increments have been extensively used to study the evolution of the magnetic field rotations in the solar wind \citep{Borovsky2010, Zhdankin2012, chenRotKin2015MNRAS.453L..64C, matteini2018ApJ, perrone2020A&A...633A.166P, wu2023ApJ...947L}.

The vector increments magnitude probability density functions (PDFs) at 1 AU computed at different time lags possess a log-normal shape \citep{Zhdankin2012} (even at kinetic scales \citep{chenRotKin2015MNRAS.453L..64C}). For each lag the parameters of the log-normal distributions are different, but the PDFs can be rescaled to a universal log-normal in the inertial range. From the universal log-normal it is possible to recover a rotation model for the magnetic rotations that fits the data well. Interestingly the distributions of the magnetic rotations at 1 AU are to a large degree reproduced by MHD turbulent simulations (if the root mean square fluctuations are of the order of or greater than the background magnetic field), suggesting that turbulence might be the leading cause of the generation of both large and small magnetic deflections in the solar wind. The results above are described in \cite{Zhdankin2012}. 

Log-normal distributions are observed in the solar wind not only for magnetic increments \citep{Zhdankin2012, chenRotKin2015MNRAS.453L..64C} but also for the magnetic field magnitude \citep{burl2001Log}, for the scale dependent energy dissipation rate \citep{zhdankin2016dissipationrate} and as probes of the energy cascade rate distributions \citep{sorrisoValvoIntermittencyCastaing1999} in the context of multiplicative random cascade models \citep{castaing1990}. 
In these models the non conservative (intermittent) behaviour of the local energy dissipation rate is modeled through the multiplication of random variables drawn from the same distribution \citep{frischbook}. 
The log-normal is one of the possible distributions choices, but it seems to be the most common one in the solar wind and it is probably a consequence of dealing with intermittent turbulent signals.
   
In order to understand the relation between turbulence and SBs we study the evolution of the magnetic increments and rotation distributions in the solar wind at different radial distances using PSP data. 
The questions we address are the following:
\begin{itemize}
   \item are the magnetic vector increments still well fitted by a log-normal function at different heliocentric distances?
   \item is there a universal log-normal as suggested by \cite{Zhdankin2012}?
   \item is the rotation model obtained at 1 AU still valid at different radial distances?
   \item do SBs arise as a separate population in these distributions?
   \item is the radial evolution of the PDFs consistent with a turbulent picture for SBs?
\end{itemize}

In Section~\ref{sec:data} we describe the data set used in this study, in Section~\ref{sec:results} we report our results and in Section~\ref{sec:conclusion} we discuss our conclusions. 
\begin{figure}
    \centering
    \includegraphics[width=\linewidth]{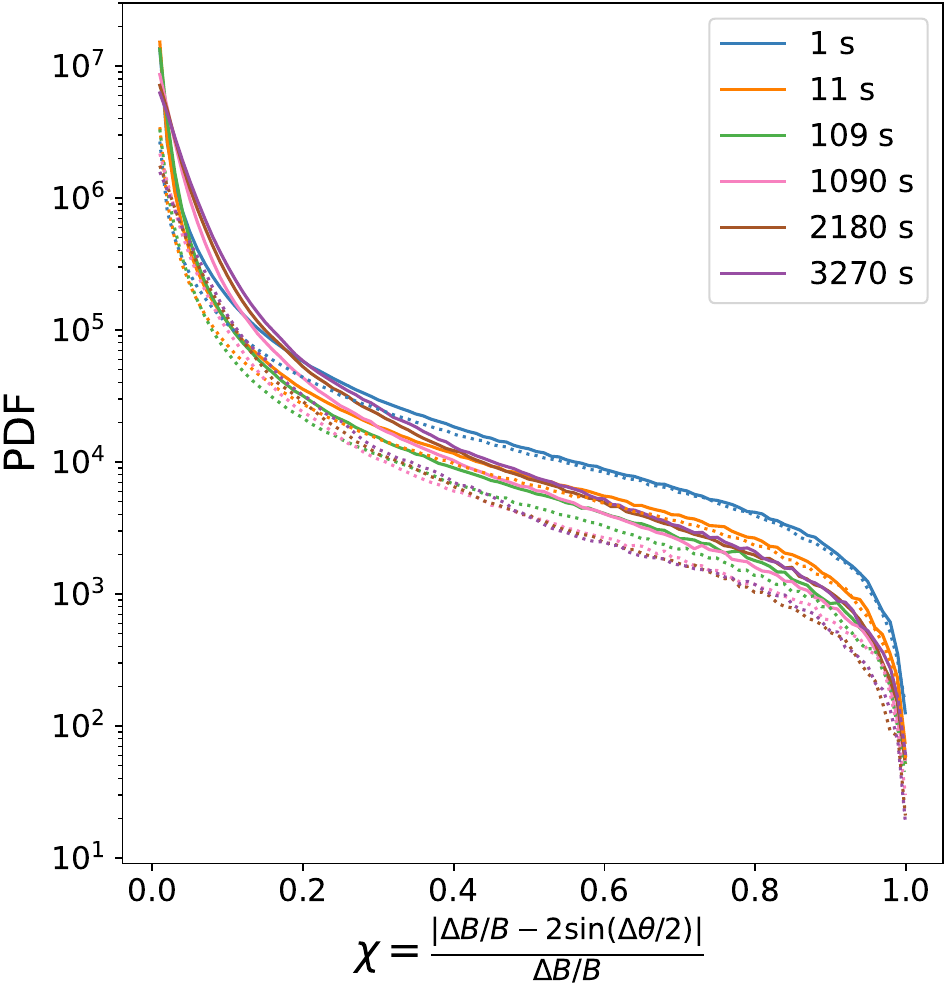}
	\caption{Distributions of $\chi$, normalized difference between the magnetic vector increments and the corresponding angle if this was due to a pure rotation. Different colors represent different values of $\tau$. Solid lines are at distances below 0.1 AU, dotted lines are at distances between 0.4 and 0.5 AU.}
	\label{fig:rotationX}
\end{figure}

\section{Data and Methods}\label{sec:data}
We use data from the fluxgate magnetometer MAG \citep{Bale2016} at 4 samples per cycle cadence and the electron pitch angle distributions (ePAD) from the SPAN-e instrument \citep{Kasper2016, phyllisSpanE2020}.
The data in this study cover the fist eleven orbits of PSP at distances below 0.5 AU. 

In the dataset transients like coronal mass ejections (CMEs) are removed by eye and the heliospheric current sheet (HCS) crossings are removed with the aid of the ePAD. CMEs are excluded because they are not part of the steady-state solar wind, the HCS crossings are removed because they are large angle rotations related to the change in polarity rather than to switchbacks or to turbulence.

We compute the distributions of the magnetic field increments: 

\begin{equation}
    \Delta B / B = \frac{\lvert \textbf{B}(t+\tau)- \textbf{B}(t) \rvert} {\lvert \textbf{B}(t) \rvert} 
\end{equation}

and the corresponding angular rotations 
\begin{equation}
    \Delta \theta = \arccos \left( \frac{\textbf{B}(t+\tau) \cdot \textbf{B}(t)} {\lvert \textbf{B}(t) \rvert \lvert \textbf{B}(t+\tau) \rvert}  \right).    
\end{equation} 

Under the assumption of pure rotations between $t+\tau$ and $t$ with no field magnitude change, the angle and the increments are related by $\Delta B / B = 2 \sin{ (\Delta \theta /2) }$ \citep{Zhdankin2012}.
Each data point in the time series provides an increment value, unless $\textbf{B}(t+\tau)$ or $\textbf{B}(t)$ are data gaps. In this case no increment value is obtained. Once we have a data series of increments at a given $\tau$ and distance we compute the corresponding distribution.

In the distributions we consider values of  $\Delta B / B$ and corresponding rotations only for increments up to 2. This upper limit is set by the fact that a 180 degrees rotation can give a maximum increment value of 2. Therefore any value larger than this cannot be the result of a pure rotation. Applying this threshold has the effect of removing part of the tail of the distributions (the part due to highly compressive increments), but less than 0.6\% of the pre-processed points (with HCS crossings and CMEs removed) are lost as a consequence. 

\section{Results}\label{sec:results}

\subsection{Evolution of the increments and rotation distributions with heliocentric distance}\label{sec:results1}

\begin{figure*}
    \centering
	\includegraphics[width=\linewidth]{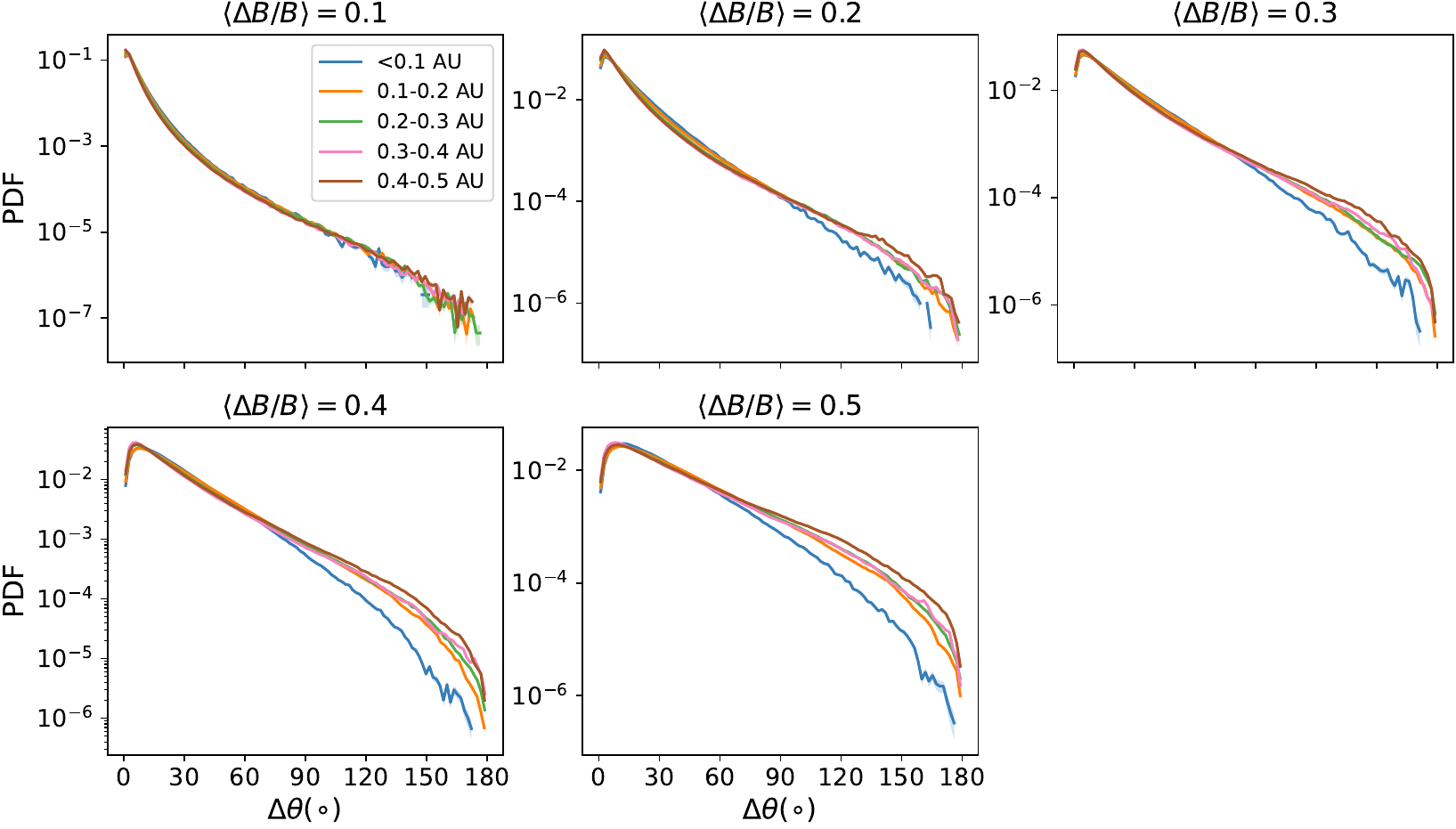} 
	\caption{Distributions of the magnetic deflection angles $\Delta \theta$ at fixed values of the average increment value $\langle \Delta B/ B \rangle$}
	\label{fig:dthetasamedB}
\end{figure*}

The evolution of the magnetic field increments with distance and scale is plotted in Figure~\ref{fig:dBovBsameTau}. The lags ($\tau$) are chosen to be within the range of residence times observed for SBs \citep{DudokdeWitSBs2020}. The different curves change position with distance with respect to one another. For small $\tau$ the closest to the Sun distribution (blue line) presents the highest occurrence of large increments, whereas it presents the lowest occurrence at large $\tau$.

In Figure~\ref{fig:dthetasameTau} the rotation distributions are shown. Not surprisingly the curves behave similar to those of Figure~\ref{fig:dBovBsameTau} since the magnetic field undergoes mostly rotation in the solar wind, especially at PSP distances. The dominance of rotations is highlighted in Figure~\ref{fig:rotationX}. The parameter $\chi$ is a measure of the deviations from pure rotations. The distributions of $\chi$ are peaked around zero independent of distance and scale with a drop of more than 2 orders of magnitude between the peak value and the value at $\chi=0.1$. This confirms the predominance of rotations in the solar wind also in the inner heliosphere.

The behavior of the PDFs in Figures~\ref{fig:dBovBsameTau} and \ref{fig:dthetasameTau} is in part due to the fact that by using the same $\tau$ at different distances we compare distributions with a different underlying average level of $\Delta B/B$ and we neglect the evolution of the $1/f$ break with distance \citep{bruno_carbone2013, chenTurb2020, ZesenHuang1ovf2023, nooshin2023}.

In order to see clearly the changes in the shape of the distributions we need to account for these effects. We do this with the following procedure: for each distance bin of Figure~\ref{fig:dBovBsameTau} we compute $\langle \Delta B/ B \rangle$ with respect to the different $\tau$, then through a linear interpolation we obtain a curve of $\langle \Delta B/ B \rangle$ against $\tau$ for each distance bin a curve. This allows us to determine a value for $\tau$, for each distance bin, that corresponds to any value of $\langle \Delta B/ B \rangle$ we chose. In this manner we obtain a different $\tau$ for each radial distance that produces the same $\langle \Delta B/ B \rangle$. The increments and the corresponding rotations are then recomputed with the new set of $\tau$.

The results are shown in Figure~\ref{fig:dthetasamedB}. 
It can be seen that the curves at small $\langle \Delta B/ B \rangle$ values share the same shape at all distances but differ at large angles for the larger $\langle \Delta B/ B \rangle$ values. 
This behavior suggests that the small scale distribution is already fully evolved at these distances while the larger scales ($\langle \Delta B/ B \rangle> 0.1$) are still in the process of evolving to their final state. Such a scale dependence is highly suggestive of a turbulence dominated evolution for the PDFs, since in a turbulent cascade the non-linear time is scale dependent with the smaller scales evolving faster.

\subsection{Log-normality and Zhdankin's rotation model with PSP}

\begin{figure*}
    \centering
	\includegraphics[width=\linewidth]{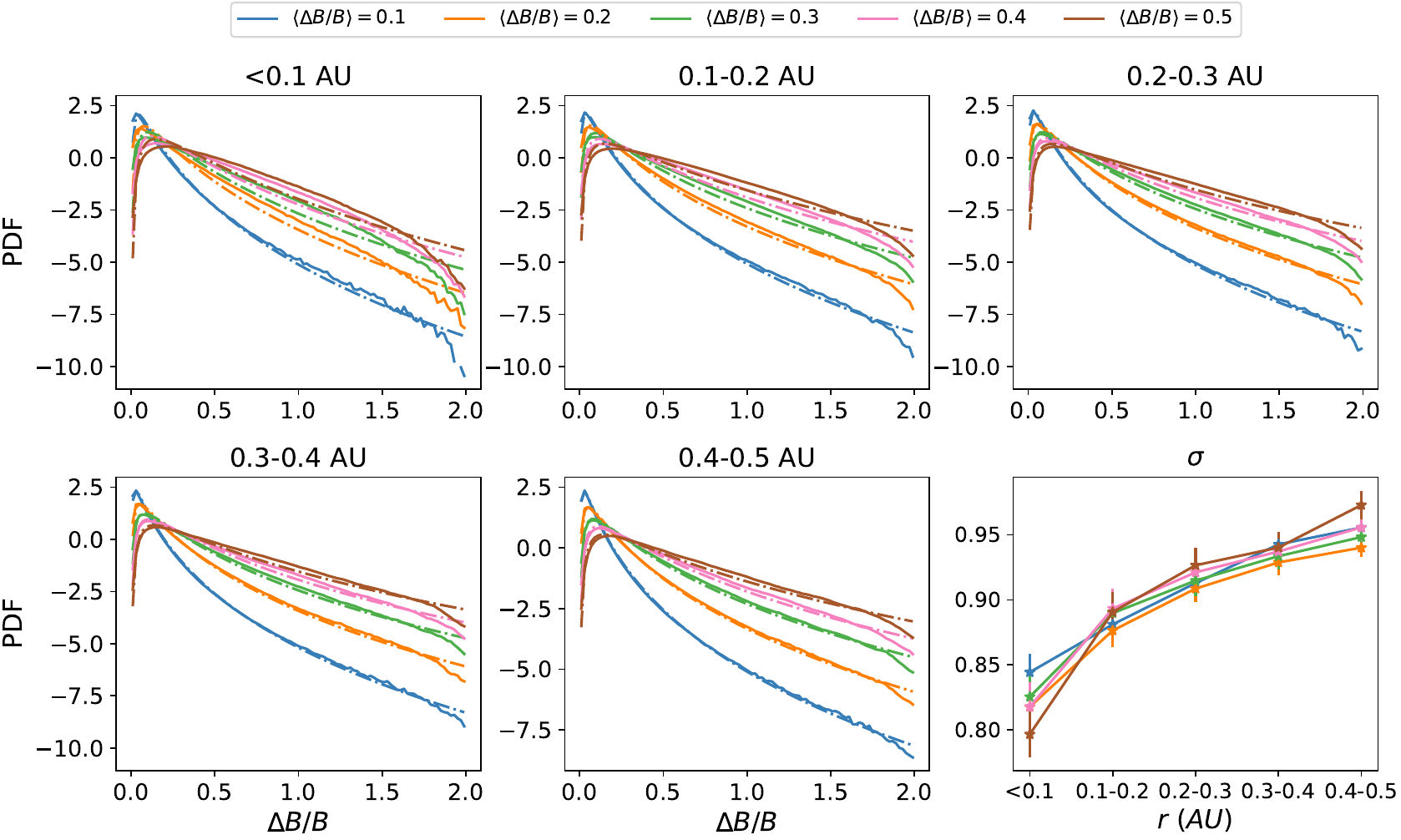} 
	\caption{Distributions of the magnetic deflection angles at fixed average $\Delta B/ B$. The dashed lines are log-normal fits to the data. The table on the lower right shows the variation with distance and scale of the $\sigma$ parameter.}
	\label{fig:fittingdB}
\end{figure*}

In Figure~\ref{fig:fittingdB} we test whether $\Delta B/ B$ follows a log-normal distribution 
throughout the full range of distances and scales considered here. The log-normal formula (Equation~\ref{lognormal}) is

\begin{equation}\label{lognormal}
\pdf    
\end{equation}
where the parameters $\mu$ and $\sigma$ represent respectively the mean and the standard deviation of the logarithm of $x$.
The results in Figure~\ref{fig:fittingdB} clearly show that as we move further out in the heliosphere the distributions are better fitted by a log-normal, even though the fit is reasonably good even at the closest distances. In order to make this statement more quantitative we compute the coefficient of determination defined as $R = 1- \Sigma_{i=1}^n (y_i - \left< y \right>)^2 / \Sigma_{i=1}^n (y_i - f_i)^2 $, where $y_i$ and $\left< y \right>$ are respectively the measured values and their mean, $f_i$ represents the values of the model and $n$ the number of data points.  
The coefficient of determination is close to one for all the curves in Figure~\ref{fig:fittingdB}. For $\langle \Delta B/ B \rangle=0.5$, the value of $R$ varies from 0.90 at distances below 0.1 AU to 0.97 at distances in the range 0.4-0.5 AU. For the smallest scales, $\langle \Delta B/ B \rangle=0.1$, the value is already 0.98 at the closest distances and reaches 0.998 for the furthest distances.
Figure~\ref{fig:fittingdB} illustrates also the radial and scale dependent evolution of the $\sigma$ parameter, which at 1 AU in Wind data is found to be $\sigma \approx 1$ \citep{Zhdankin2012}. We observe that $\sigma$ increases with increasing radial distance for all values of $\langle \Delta B/ B \rangle$. At the closest distances the distributions with $\langle \Delta B/ B \rangle = 0.1$ possess the closest value to 1. 
We also test whether other functions commonly used in the literature to describe the magnetic increments fit them as good as the log-normal function. A double exponential was used by \cite{Borovsky2010} to fit $\Delta \theta$, but it can be tested also for $\Delta B/ B$.  This function, that has two more fitting parameters than the log-normal, gives a coefficient of determination very close to one only for $\langle \Delta B/ B \rangle$ up to 0.3 but at larger values the fits fail to converge. This is due to the impossibility of reproducing the low $\Delta B/ B$, downward section of the curve at $\langle \Delta B/ B \rangle > 0.3$ with a double exponential, the same problem is found when fitting $\Delta \theta$.
We also test the log-Poisson distribution which is observed for other measures of turbulence \citep{zhdankin2016dissipationrate}, but the fits in this case give a poor agreement (not shown). The log-normal seems to be the strongest candidate distribution for the solar wind fluctuations.
The log-normality of the magnetic vector increments can be linked to turbulence in the context of random cascade models and might be ultimately linked to the log-normality of the scale dependent dissipation rate \citep{zhdankin2016dissipationrate}.

\begin{figure*}
    \centering
	\includegraphics[width=\linewidth]{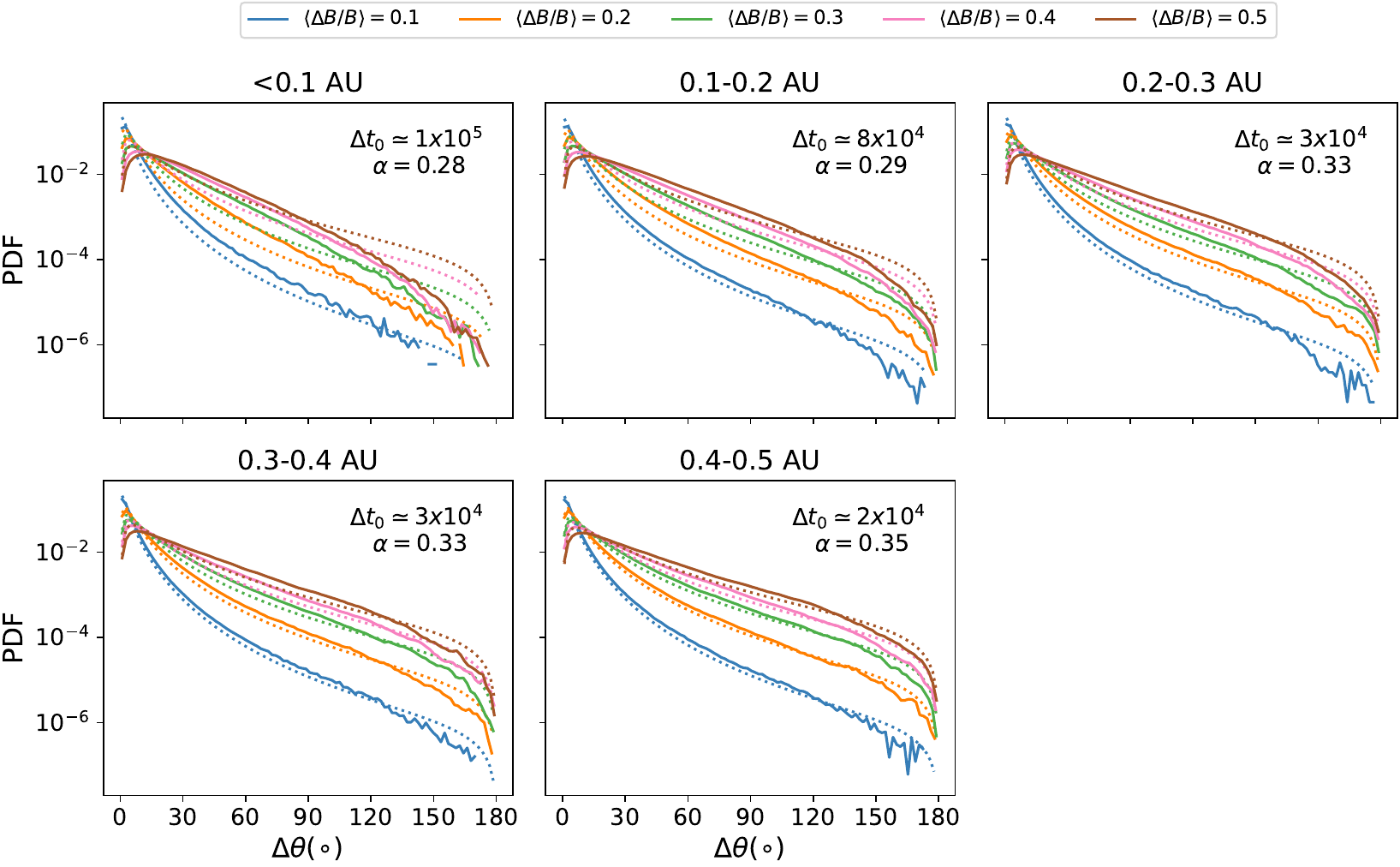} 
	\caption{Distributions of $\Delta \theta$. The dotted lines represent the model of Equation~\ref{eq:rotmodel}. The fitting parameters are indicated in each panel and the distances on top of each panel.}
	\label{fig:fittingdtheta}
\end{figure*}

In Figure~\ref{fig:fittingdtheta} we compare the PSP rotation distributions with the rotation model (Equation~\ref{eq:rotmodel}) developed by \cite{Zhdankin2012}, 

\begin{equation}\label{eq:rotmodel}
\begin{aligned}        
     & g(\Delta \theta) = \frac{1}{K\sqrt{8\pi} \tan{\frac{\Delta \theta}{2}}} \times  \\ 
     &\exp{ \left( -\frac{1}{2} \log^2{ \left( 2\sin{ \left(\frac{\Delta \theta}{2} \right)} \left( \frac{\tau}{\Delta t_0}\right) ^{-\alpha}  \right) }\right)} 
\end{aligned}
\end{equation}
where $\Delta t_0$ and $\alpha$ are fitting parameters and \\ 
 $K = \frac{1}{2}\erf{\left[\log{\left (2\left( \frac{\tau}{\Delta t_0}\right) ^{-\alpha}\right)/\sqrt{2} +1}\right]} $ is a normalization constant (independent of $\Delta \theta$). At 1 AU $\Delta t_0$ is interpreted to be the outer scale of the turbulence \citep{Zhdankin2012}. In Figure~\ref{fig:fittingdtheta} a different $\tau$ is used for each $\langle \Delta B/ B \rangle$ and for each distance, as discussed in Section~\ref{sec:results1}.
The key ingredients of the model are the possibility to rescale the increments PDFs for different $\tau$ into a single log-normal given that $ \sigma \approx 1$, which is the case at 1 AU in the inertial range, and the fact that the increments are assumed to be due to pure rotations.

The agreement between the PSP distributions and the Zhdankin rotation model (Equation~\ref{eq:rotmodel}) improves with increasing radial distance as shown in Figure~\ref{fig:fittingdtheta}, but at $\langle \Delta B/ B \rangle=0.1$ the agreement is quite good even close to the Sun. This is consistent with the evolution of the increment distributions, since for $\langle \Delta B/ B \rangle=0.1$ the log-normal fit gives a coefficient of determination closer to one than in the other cases. 
The reason why, for $r<0.1 \ AU$, the fit for larger $\langle \Delta B/ B \rangle$ is not as good does not have to be attributed to the transition to the $1/f$ range. In fact, even though for $\langle \Delta B/ B \rangle = 0.5$ the lag $\tau \simeq 7 \times 10^2s$ is in the $1/f$ range, the fits for $\langle \Delta B/ B \rangle = 0.3$, whose $\tau \simeq 50s$ is in the inertial range show the same behaviour.
We attribute then, the discrepancy between the model and the observations in Figure~\ref{fig:fittingdtheta} to the fact that the distributions of the magnetic increments below 0.1 AU are not yet fully evolved to the log-normal with $\sigma=1$, i.e. the universal log-normal proposed by \cite{Zhdankin2012}.
Consistent with the evolving state of the distributions, the fitting parameters $\Delta t_0$ and $\alpha$ with increasing radial distance get closer to the value of $\Delta t_0 =6600$ and $\alpha =0.46$ observed at 1AU \citep{Zhdankin2012}.    

\section{Discussion and conclusions}\label{sec:conclusion}

We presented the first report on the radial evolution of the scale-dependent increments and rotation angle distributions for distances below 0.3 AU.

Our results show that the rotation angle distributions (Figure~\ref{fig:dthetasameTau} and Figure~\ref{fig:dthetasamedB}) 
evolve with radial distance in a scale dependent fashion.
In agreement with this, the increment distributions are still evolving towards log-normality but this evolution is different for the PDFs at small and large values of $\langle \Delta B/ B \rangle$. At $\langle \Delta B/ B \rangle=0.1$ i.e. small scales, the coefficient of determination is close to one at all distances while for larger values i.e. large scales, it is still evolving with distance towards one (Figure~\ref{fig:fittingdB}).
This suggests a scale dependent evolution towards a log-normal shape, with the small scales being approximately log-normal independent of the distance. 
The log-normal though is not the universal one proposed in \cite{Zhdankin2012} because $\sigma$ is not equal one at the distances investigated here, but $\sigma$ does evolve towards one with increasing radial distance. 

The evolution of $\sigma$ for the small scales seems somewhat contradictory with the distributions having the same shape in Figure~\ref{fig:dthetasamedB}. The behaviour of $\sigma$ though is dominated by the tails of the distributions since we are fitting in log-space. The reason for the similar behavior between small and large scales is that even at the smallest scales there is some evolution in the far tail of the distributions, it is possible to see this evolution in the $\langle \Delta B/ B \rangle=0.1$ curves in Figure \ref{fig:fittingdB}.

In the rotation distributions switchbacks do not arise as a distinct population, in the sense that they do not appear as an extra bump at large angles. Furthermore, as illustrated in Figure~\ref{fig:fittingdtheta}, a single function (\ref{eq:rotmodel}) based on the log-normality of the increments is capable of capturing most of the rotations, other than those at close distances and large scales, where there are fewer large angle rotations. This suggests that switchbacks, considered as large-angle rotations, are part of a single distribution of solar wind fluctuations, as might arise, for example, from a turbulent cascade.
The results shown here support the in-situ (during propagation in the heliosphere, not right at the spacecraft) formation of switchbacks. In fact, the large-scale PDFs at large angles, where most of the large angle deflections are present (see Figure~\ref{fig:dthetasameTau}), are increasingly filled with increasing radial distance indicating the presence of more switchbacks, in agreement with the results of \cite{pecoraSBs2022ApJ, Jagarlamudi_2023}. This behaviour is not expected from the ex-situ models unless combined with a shear or turbulence/Alfv\'en wave based mechanism. 

The scale dependent evolution towards a log-normal shape and the change in shape of the the PDFs even at fixed $\langle \Delta B/ B \rangle$ is a key property to consider to investigate the origin of the distributions.
The change in shape has two possible interpretations. \textit{One} The turbulent interactions in the solar wind are reshaping the distribution into a log-normal. Indeed turbulence simulations are able to approximately produce log-normal distributions for the magnetic field vector increments, and can reproduce the rotation distributions at 1 AU \citep{Zhdankin2012}. Furthermore log-normal distributions are observed in turbulence simulations for the scale dependent energy dissipation rate and in solar wind data for a proxy of the same quantity \citep{zhdankin2016dissipationrate}.
The scale dependent evolution of the distributions is also consistent with a turbulence scenario. Turbulence interactions are faster at smaller scales, so one would expect the larger scales to evolve more slowly, in agreement with our results.

\textit{Two} The change in shape could be attributed to the growth of the fluctuations with the expansion with the constraint of having a constant magnetic field magnitude.
This constraint has to be invoked because expansion alone can grow the amplitudes of $ \Delta B/ B$ \citep{Parker1965SSRv....4..666P, belcher1971WKB, MalletSBs2021, squire_mallet_3dAlfvSol_2022}, i.e., shift the unnormalised PDFs to larger $ \Delta B/ B$ values, resulting in a growth of large angular deflection, but is not expected to change the shape of the PDFs (see Figure~\ref{fig:dthetasamedB}), therefore an additional process is required to explain the full distribution of rotations, including the switchbacks.
At large scales, this constraint implies that there is a cutoff to the distribution at $ \Delta B/ B=2$, as a consequence the PDF perhaps changes its shape once this cutoff is reached due to the expansion driven growth of the fluctuations. However, it is not clear why such a cutoff would cause the PDFs to become log-normal, and it would not explain why the PDFs are log-normal at small scales.
Furthermore, the physical origin of the constraint is also an open question \citep{barnesHollweg1974, vasquez_hollweg1998, roberts2012prl, matteini2015ApJ, tenerani_velli2018AGUFMSH53A..06T, squire2019JPlPh..85a9014S, squire_mallet_3dAlfvSol_2022}. 

Considering the results shown and the considerations made here, it seems most likely that expansion is causing the overall amplitudes to grow, and turbulence is reshaping the magnetic field rotations to create the fluctuation distributions that we measure.

\section*{Acknowledgements}
 AL is supported by STFC Consolidated Grant ST/T00018X/1. CHKC is supported by UKRI Future Leaders Fellowship MR/W007657/1 and STFC Consolidated Grants ST/T00018X/1 and ST/X000974/1. JRM is supported by STFC studentship grant ST/V506989/1. VKJ acknowledges support from the Parker Solar Probe mission as part of NASA's Living with a Star (LWS) program under contract NNN06AA01C. JRM and AL acknowledge support from the Perren Exchange Programme. We thank the
members of the FIELDS/SWEAP teams and PSP community
for helpful discussions.

\section*{Data Availability}
PSP data are available at the SPDF
(https://spdf.gsfc.nasa.gov).

\bibliographystyle{mnras}

\bsp	
\label{lastpage}
\end{document}